\documentclass[10pt,english]{IEEEtran}
\usepackage[T1]{fontenc}
\IEEEoverridecommandlockouts
\usepackage{geometry}
\usepackage{amsmath,amssymb,amsfonts}
\usepackage[linesnumbered,ruled]{algorithm2e}
\usepackage[caption = false]{subfig}
\usepackage{graphicx}
\usepackage{textcomp}
\usepackage[dvipsnames]{xcolor}
\usepackage{colortbl}

\usepackage[colorlinks=true, allcolors=blue]{hyperref}
\usepackage{float}

\title{XAI-Driven Client Selection for Federated Learning in Scalable 6G Network Slicing}

\author{
    \IEEEauthorblockN{Martino Chiarani\IEEEauthorrefmark{1}, Swastika Roy\IEEEauthorrefmark{2}, Christos Verikoukis\IEEEauthorrefmark{3}, Fabrizio Granelli\IEEEauthorrefmark{1}}  
    \IEEEauthorblockA{\IEEEauthorrefmark{1} University of Trento, Italy}  
    \IEEEauthorblockA{\IEEEauthorrefmark{2} Iquadrat Informatica S.L., Barcelona, Spain}  
    \IEEEauthorblockA{\IEEEauthorrefmark{3} University of Patras, Greece}  
    \IEEEauthorblockA{Emails: martino.chiarani@studenti.unitn.it, s.roy@iquadrat.com, cveri@ceid.upatras.gr, fabrizio.granelli@unitn.it}  
}

\begin{document}

\maketitle

\begin{abstract}

In recent years, network slicing has embraced artificial intelligence (AI) models to manage the growing complexity of communication networks. In such a situation, AI-driven zero-touch network automation should present a high degree of flexibility and viability, especially when deployed in live production networks. However, centralized controllers suffer from high data communication overhead due to the vast amount of user data, and most network slices are reluctant to share private data. In federated learning systems, selecting trustworthy clients to participate in training is critical for ensuring system performance and reliability. The present paper proposes a new approach to client selection by leveraging an XAI method to guarantee scalable and fast operation of federated learning based analytic engines that implement slice-level resource provisioning at the RAN-Edge in a non-IID scenario. Attributions from XAI are used to guide the selection of devices participating in training. This approach enhances network trustworthiness for users and addresses the black-box nature of neural network models. The simulations conducted outperformed the standard approach in terms of both convergence time and computational cost, while also demonstrating high scalability.
\\
\end{abstract}


\vspace{9mm}

\section{Introduction}
6G networks promise to uphold multiple applications with highly diverse performance requirements, making resource allocation optimization critical for meeting these demands \cite{5g_slice_survey}. In particular, automated network slicing through AI is one of the most encouraging technologies, addressing this challenge by implementing the zero-touch network and service management (ZSM) framework standardized by ETSI in \cite{ETSI_ZSM}.
Nevertheless, with the increasing number of connected devices, collecting and processing all of this data centrally becomes a constraint of both communication and computation tasks, while also raising privacy concerns. Accordingly, Federated Learning (FL) offers a solution by decentralizing the learning process, allowing individual devices to train AI models locally without sharing raw data, thus avoiding large-scale data collection \cite{net_slic_FL_approach}.

However, it is important to consider that with 6G, devices range from simple IoT sensors to smartphones, each generating different types of data. For this reason, standard FL models may suffer from convergence delays and higher computational costs if trained on non-IID data \cite{FL_prob}. Additionally, the potential for an innumerable number of clients could push the FL capability beyond its limits, as it must communicate the locally trained model updates back to a central server, creating a vast communication bottleneck. Therefore, efficient client selection \cite{CL_in_FL_princ} that reduces unnecessary communications can improve model accuracy \cite{eff_FL_part_select}, reduce training overheads \cite{CL_frame_eff_FL}, and strengthen robustness \cite{Fast_conv_FL}. Moreover, the fact that a fully automated network will be completely detached from human collaboration places more emphasis on the ability to interpret the models’ decisions. The black-box behavior of AI models can make it difficult to understand how decisions are made, which can lead to a lack of trust between clients and operators \cite{AI_interpret_myth}.

Hence, to tackle the challenges of non-IID datasets and build a more trustworthy global model, a promising approach is to use eXplainable AI (XAI) techniques, such as Integrated Gradients \cite{Axiomatic}, as applied in this case. XAI methods help weigh the contributions of individual features to a model's predictions by attributing importance scores to input features. In the context of FL, where data remain decentralized and heterogeneous, explainability plays a critical role in improving transparency and trust in the system \cite{Fed_XAI}.
By leveraging XAI methods in FL, operators obtain a clearer understanding of why a certain model output is produced, which is also crucial for client selection. Traditional FL algorithms struggle when client data distributions are non-IID because clients may have different feature distributions or data quality \cite{FL_nonIID_data}. XAI can address this issue by evaluating the importance of the features used by individual clients, offering a way to prioritize clients whose data return more significant contributions to the global model.
This process allows the selection of clients with more valuable datasets, ensuring that clients with more impactful and informative data are chosen more frequently \cite{CL_FL_pow}. This approach aligns with the idea of selection clients based on the importance of their contributions, rather than the uniform or random selection of clients \cite{CL_FL_review}. 
In this way, the global model will be trained on the most relevant and high-quality data, which not only improves its performance but also helps mitigate the biases introduced by non-IID datasets \cite{XAI_FL_fraud_detect}. Thus, the global model is trained on relevant data with respect to the features.

\subsection{Related Work}

As discussed in \cite{net_slic_FL_approach} and \cite{MEC_FL_slice}, FL in network slices is considered an adequate solution for predicting service-oriented Key Performance Indicators (KPIs) of slices. It achieves good prediction accuracy, can dynamically allocate resources for multiple slices, and provides the corresponding QoS requirements. In \cite{FedAVg}, the focus was on gradient-descent-based FL (FedAvg), which was designed to address the non-IIDness of local datasets and analyze the convergence bound. 
In addition, an SLA-driven stochastic policy in a non-IID setup was proposed in \cite{SLA_FL_net_slic} to ensure scalability and rapid operation during FL resource provisioning.
The survey \cite{CL_FL_Survey} provides a comprehensive overview of client selection techniques in FL, including their strengths and limitations, as well as the challenges and open issues that need to be addressed.
Furthermore, to improve performance, a valid client selection method was presented in \cite{CL_grad_import} for enhancing the communication efficiency of FL systems by selecting devices based on the norms of their gradient values. Additionally, the client selection through an XAI approach in FL was tested in vehicular scenarios by \cite{Vehic_XAI_SHAP} and \cite{Vehic_XAI_LIME}, where the results seem promising. The XAI methods used are SHAP values and LIME, respectively, which are employed to generate a trust value, or score, to select the most valuable clients for training.
In the network automation scenario from \cite{Explain_FDL_slic}, XAI is used together with FL to predict the probability of a traffic drop, whereas sensitivity- and explainability-aware metrics are considered constraints.

\subsection{Contributions}
This paper presents the IntelliSelect-FL approach, which predicts CPU usage and leverages attributions from XAI methods for Analytic Engine (AE) selection during federated learning training. This ensures transparent Zero-Touch Service Management (ZSM) of 6G slice RAN in non-IID conditions.
In detail,
\begin{itemize}
\item To handle the FL training task for local AEs, the Integrated Gradients XAI method has been implemented for feature attribution calculations, which are used to calibrate the importance of each feature with respect to the AEs. 
\item During each iteration, the client selection phase is executed by evaluating the combination of specific attributions of all participants. This approach ensures a transparent and well-informed decision on which AEs are best suited to contribute to the next round of FL.
\item To guarantee scalability in the slicing context, an XAI-driven FL policy has been designed to select a subset of AEs that will participate in the FL task at each iteration round, improving the convergence time while maintaining the same performance, regardless of how many AEs are added to the network.
\end{itemize}

\begin{figure}
\centering
\includegraphics[scale=0.64]{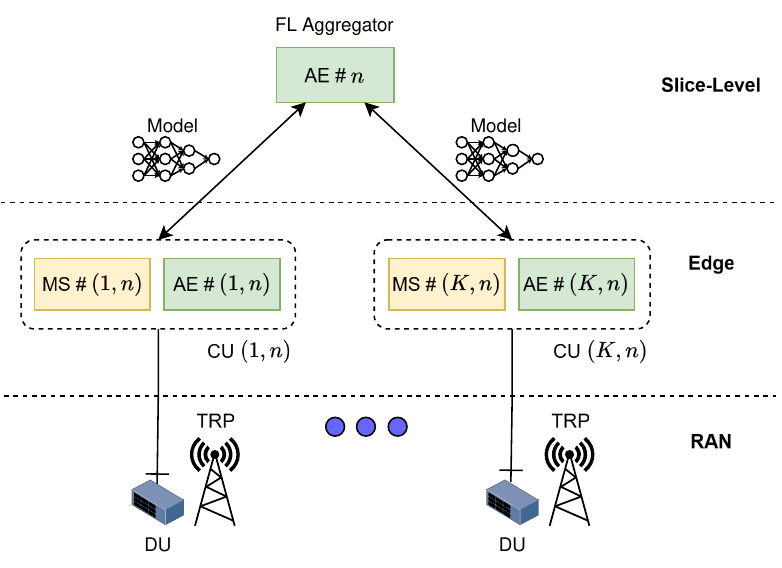}
\caption{\label{fig:Archi_prop} Proposed architecture.}
\end{figure}

\begin{table}
\centering
\caption{Features and Outputs Description}
\begin{tabular}{c c}
\hline
\rowcolor{lightgray}
Feature  & Description  \\ 
\hline
   & Apple, Facebook, Facebook Messages, \\
OTT Traffics & Facebook Video, HTTPS, Instagram,  \\
 & Netflix, QUIC, Whatsapp, Youtube   \\
\\
CQI  & Channel quality indicator  \\
\\
MIMO-FI & MIMO rank usage (\%)  \\ 
\hline 
\rowcolor{lightgray}
Output & Description \\ 
\hline
CPU Load & CPU resource consumption (\%)  \\ \hline\hline
\end{tabular}
\label{table:Feature desc}
\end{table}

\section{Proposed Solution}
The architecture of the proposed solution is depicted in Fig. \ref{fig:Archi_prop}, which presents a 6G RAN topology consisting of a per-slice central unit (CU) and distributed unit (DU) functional split, with the latter element associated with a transmission/reception point (TRP); this configuration is inspired by this article \cite{SLA_FL_net_slic}.
Each CU $k$ $(k=1,...,K)$ is composed of a monitoring system (MS) and an AI-enabled slice resource allocation function, named Analytic Engine (AE). Each AE performs slice-level RAN KPI data collection to create local datasets for slice $n$ $(n=1,...,N)$, i.e., $D_{k,n}=\{x^{(i)}_{k,n}, y^{(i)}_{k,n}\}^{D_{k,n}}_{i=1}$.


\begin{figure*}[t!]
\centering
\includegraphics[scale=0.8]{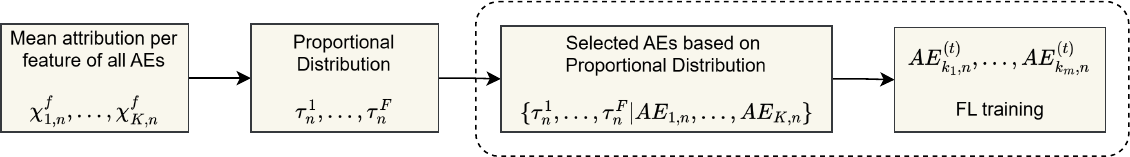}
\caption{\label{fig:Prop_pol_cl} Proposed policy for AE selection.}
\end{figure*}

The considered datasets, as shown in Table \ref{table:Feature desc}, are collected using Tektronix probes and the Huawei PRS OSS tool from 3200 TRPs in a real LTE-advanced network \cite{dataset-4g}.  Each dataset has 1000 data points with a granularity of one hour. These TRPs encompass a range of regions, including residential, business, and event zones, that display different traffic patterns due to heterogeneous user distributions and profiles.
Every dataset's Input features include hourly traffic from major over-the-top (OTT) applications, CQI, and MIMO full-rank usage. The output KPI includes the CPU load \cite{dataset-4g}. 
It should be noted that, although the dataset originates from a 4G system, the core challenge of this work is to address AI-driven resource management, which remains critical for 6G networks. The considered dataset reflects realistic traffic dynamics, non-IID conditions, and time-varying channel effects, closely aligned with the challenges of FL-based orchestration in 6G.
Direct validation on a live 6G network is not yet feasible due to the lack of publicly available data. However, this methodology can be extended to future 6G deployments, presenting a potential direction for future research.

Moreover, these datasets are non-IID due to the varying traffic profiles resulting from heterogeneous user distributions and channel conditions, as previously mentioned, which adds complexity to the training of FL algorithms.

Existing approaches to addressing non-IIDness in FL, such as local fine-tuning and data sharing, have been shown to enhance convergence performance \cite{non_IID_Survey}. Among these, algorithm-based with local fine-tuning strategies, like FedAvg, aim to balance local and global knowledge to improve model generalization. However, while these methods help mitigate data heterogeneity, they often do not consider client selection inefficiencies. In contrast, the proposed approach not only utilizes local fine-tuning through FedAvg but also incorporates an intelligent client selection mechanism. By prioritizing the most valuable AEs, this method ensures a more robust and efficient learning process, leading to improved convergence performance.

Our presented solution involves utilizing feature importance explanations from an XAI method to understand which subset of AEs is most suitable to improve the performance of the global model in each round.
In the initialization phase, after distributing the weights of the starting global model to all AEs, the Integrated Gradients method is used by the local AEs to obtain the attribution scores.
Providing these attributions helps to understand complex models and how input features affect the neural network's outcome.
These explanations are generated by producing attribution scores for each feature of a subset of $I$ samples, using the formula from \cite{Axiomatic}:

\begin{gather} \label{eq:IG}
\scalebox{0.95}{$
IG_{k,f}^{(i)}(x)::= (x_i-x_i^{'})\times\int_{\alpha=0}^{1}{\frac{\partial{F(x^{'}+\alpha\times(x-x^{'}))}}{\partial{x_i}}}
$}
\end{gather}
The baseline, $x^{'}$, considered is the zero vector.

After obtaining all these values for each sample of each AE, their absolute mean provides the general explanation for predictions of a single AE.

\begin{gather}
IG'_{k,f}=\frac{1}{I}\sum_{i=1}^{I} \left|IG_{k,f}^{(i)}\right|
\end{gather}

Then, for each AE, these averaged attributions are proportionally normalized using the following formula:

\begin{gather}
\chi_{k,f}=\frac{IG'_{k,f}}{\sum_{j=1}^{F} IG'_{k,j}}
\end{gather}
So that, the normalized values will sum up to 1 representing the degree of relevance of each feature relative to the total. In this operation, each AE independently assesses the importance of each feature in the local data it processes. This evaluation produces an array of weights, where each element in the array corresponds to a specific feature and indicates its significance according to that particular AE.

Subsequently, they upload the information to the central server. It collects these arrays from all AEs and combines them by computing the absolute mean of the weights for each feature across all participants.
\begin{gather}
\tau_{f}=\frac{1}{K} \sum_{k=1}^{K} \left|\chi_{k,f}\right|
\end{gather}
This averaging creates a unified estimate of feature importance, incorporating insights from all participating AEs, making the selection possible.

Next, the main loop begins. Client selection is the process of choosing a subset of AEs based on the proportionality of the relevance of the features, estimated from the attributions of the Integrated Gradients. The procedure is executed independently for all slices.
This solution can be seen as a particular case of apportionment, where AEs with the highest attribution score are selected according to the proportional distribution of feature importance, as presented in Fig. \ref{fig:Prop_pol_cl}.

Once the selection is made, the selected clients conduct the training on local datasets, updating their weights for $L$ epochs, using Adam optimizer \cite{Adam}.

Finally, model averaging for each round is performed on the server following the FedAvg formula: 
\begin{gather}
W_{n}^{(t+1)}=\sum_{k\in\{k_1,...,k_m\}} \frac{D_{k,n}}{D_n}W_{k,n}^{(t)}
\end{gather}
In each round, the updated global model is distributed to all the AEs, allowing them to update their local models. This enables the computation of new local explanations, which will be collected and averaged by the aggregator, so that new proportions of the importance of the features will be used for the client selection phase in the following round.

This entire procedure is presented in Algorithm \ref{alg:PropSol}.


\SetKwInput{KwInput}{Input}                
\SetKwInput{KwOutput}{Output}
\SetKwFunction{FClientSelectionPolicy}{Client\_Selection\_Policy}

\begin{algorithm}
\caption{IntelliSelect Federated Learning Policy}
\label{alg:PropSol}
\KwInput{\textit{K, m, $S_{d}$, T, F, L}}

Set the $seed=S_{d}$\\
Initialize the global NN model \\
Distribute the global model to all the local participants

\For{k=0,...K-1}{
    Calculate IG attributions from Eq. \eqref{eq:IG} of $I$ samples, using NN local model and dataset

    \For{f=0,...,F-1}{
        $IG'_{k,f}=\frac{1}{I}\sum_{i=1}^{I} \left|IG_{k,f}^{(i)}\right|$ \textit{\textcolor{Green}{\#Average IG}
        }

    }
}
$\chi_{k,f}=\frac{IG'_{k,f}}{\sum_{j=1}^{F} IG'_{k,j}}$ \textit{\textcolor{Green}{\#Attribution normalization}}

\For{t=0,...T-1}{
    
    \scalebox{0.9}{\texttt{$\{k_{1},...,k_{m}\} =$ Client\_Selection\_Policy($m,\ \chi_{k,f}$)}}

    \For{$k \in \{k_{1},...,k_{m}\}$}{
       Train the local NN model  for $L$ epochs, respecting the Adam optimizer       
    }
    
    $W_{n}^{(t+1)}=\sum_{k\in\{k_1,...,k_m\}} \frac{D_{k,n}}{D_n}W_{k,n}^{(t)}$ \textit{\textcolor{Green}{\#FedAvg}}

    \For{k=0,...K-1}{
    Calculate IG attributions from Eq. \eqref{eq:IG} of $I$ samples, using NN local model and dataset

        \For{f=0,...,F-1}{
        \scalebox{1}{$IG'_{k,f}=\frac{1}{I}\sum_{i=1}^{I} \left|IG_{k,f}^{(i)}\right|$}
        }

    }
    
    $\chi_{k,f}=\frac{IG'_{k,f}}{\sum_{j=1}^{F} IG'_{k,j}}$ \textit{\textcolor{Green}{\#Attribution normalization}}\\
    Evaluate the MSE of global NN model using test set
    
}

\KwOutput{\scalebox{1}{MSE\_per\_round, Computational\_time\_per\_round}}

\SetKwProg{Fn}{Function}{:}{}
  \Fn{\FClientSelectionPolicy{$m,\ \chi_{k,f}$}}{
        
        $\tau_{f}=\frac{1}{K} \sum_{k=1}^{K} \chi_{k,f}$ \textit{\textcolor{Green}{\#Features average}}

        Compute the number of candidates to select for each feature
        
        Rank candidates for each feature

        Select $m$ candidates proportionally to each feature importance
        
        \KwRet indices of the selected AEs
  }
\end{algorithm}

\section{Results}

\subsection{Parameters Setting}
To investigate the proposed FL policy, three main slices are defined as stated below:
\begin{itemize}
\item eMBB: Netflix, Youtube and Facebook Video;
\item Social Media: Facebook, Facebook Messages, WhatsApp and Instagram;
\item Browsing: Apple, HTTP and QUIC.
\end{itemize}

Once network slices are defined, the traffic from the underlying OTT applications is aggregated to determine the traffic per slice, as previously mentioned. For more information on the data retrieval of the slices, refer to \cite{slices}.



This work implements a customized artificial neural network (ANN) for local training and an XAI approach to calculate feature attributions using TensorFlow \cite{tensorFlow} within a Python environment. It comprises 3 neurons as input layer (one for each feature); 2 hidden layers with  3 and 2 neurons, respectively; and 1 output neuron as output layer. The ANN’s neurons are sequentially fully connected using ReLU as the activation function and Adam as the optimization method. The parameters settings are listed in Table \ref{table:tab_param}.

\begin{figure}[t!]
    \centering
    \subfloat[ \centering Proposed Solution vs. no-policy]{
    \label{fig:Sol_MSE}
          \includegraphics[scale=0.5]{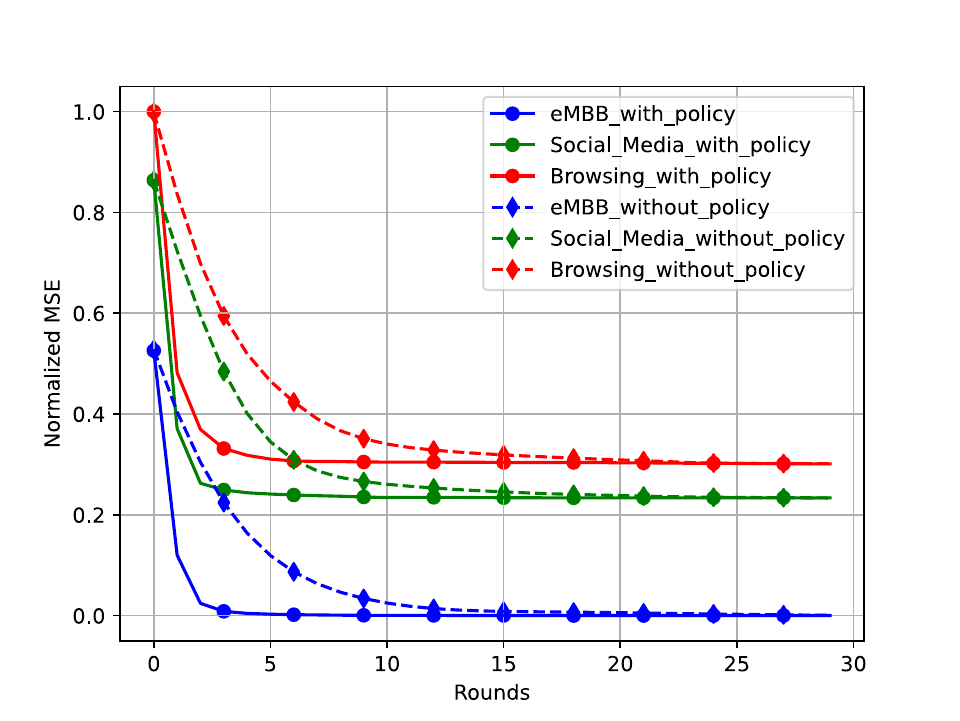}}
    
    \subfloat[ \centering Proposed Solution vs. Score Solution]{
    \label{fig:Sol_MSE_score}
          \includegraphics[scale=0.5]{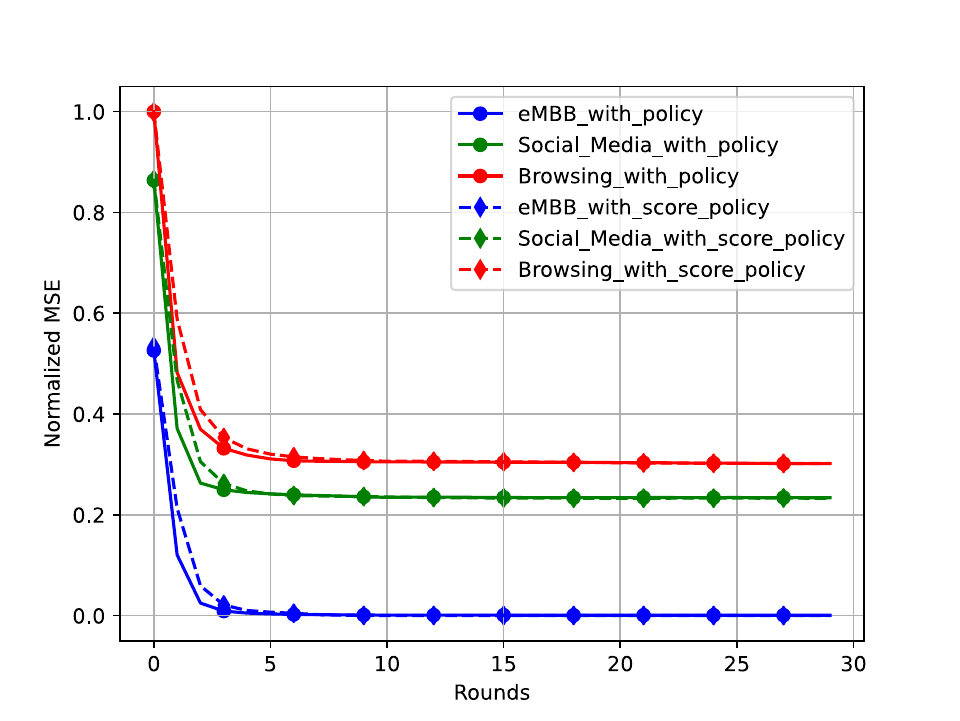}}
\caption{FL training MSE loss vs. number of FL rounds for $m=5$ and $K=10$.}
\label{MSE}
\end{figure}

\begin{table}
\caption{Settings} 
\centering 
\begin{tabular}{c c c} 
\hline 
\rowcolor{lightgray}
Parameters & Description & Value \\ [0.5ex] 
\hline 
\textit{N} & \# Slices & 3 \\ 
\textit{F} & \# Input features & 3 \\
\textit{K} & \# Total AEs & 10, (40, 50) \\
\textit{m} & \# Selected AEs & 5, 25 \\
\textit{$D_{k,n}$} & Local dataset size & 1000 samples \\
\textit{$I$} & Attribution samples & 150 samples \\
\textit{T} & \# Max FL rounds & 30 \\ 
\textit{L} & \# Local epochs & 150 \\
\textit{$\eta_{\lambda}$} & Learning rate & 0.0015 \\
\textit{$S_{d}$} & Seed & 42 \\ [1ex] 
\hline\hline 
\end{tabular}
\label{table:tab_param} 
\end{table}

\subsection{Simulated Results}

\subsubsection{Baselines Description}
\paragraph{No-policy}

The first comparison is between the proposed solution and an FL implementation with no client selection policy. This means that after initialization, where the server distributes the global weights, all clients are involved in every training round without any type of client selection or XAI explanation. In each iteration, FedAvg is still used as the aggregation method after training.

\paragraph{Score Solution}
A second comparison is made with the paper \cite{Vehic_XAI_LIME}. In that paper, the focus is also on the selection of a subset of clients (autonomous vehicles, AVs) using the XAI explanations. However, the approach is different from the one presented here because, in that case, a score (hence the name "Score Solution") is assigned to each AV, and the central server chooses the clients with the highest score. Regarding the XAI technique, the authors use LIME, but it is much slower than the Integrated Gradients method. So, for a fair comparison, the tests in this study are made utilizing the Integrated Gradients as well.

\subsubsection{Comparison Results}
\begin{figure}[htbp]
    \centering
    \subfloat[ \centering Proposed Solution vs. no-policy]{
    \label{fig:Sol_Time}
          \includegraphics[scale=0.5]{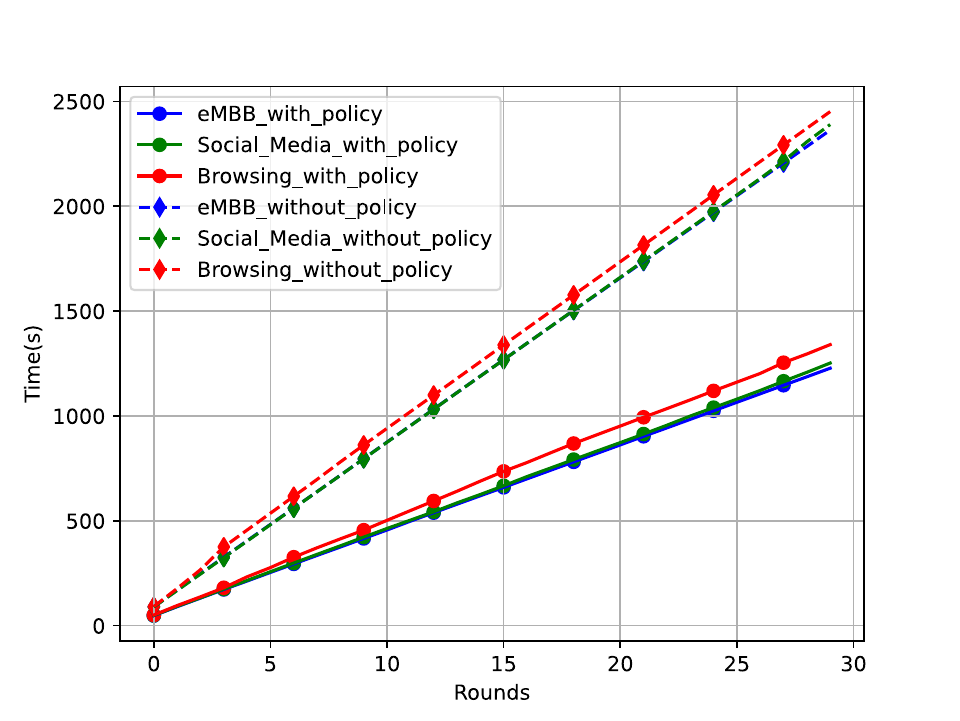}}
    
    \subfloat[ \centering Proposed Solution vs. Score Solution]{
    \label{fig:Sol_Time_score}
          \includegraphics[scale=0.5]{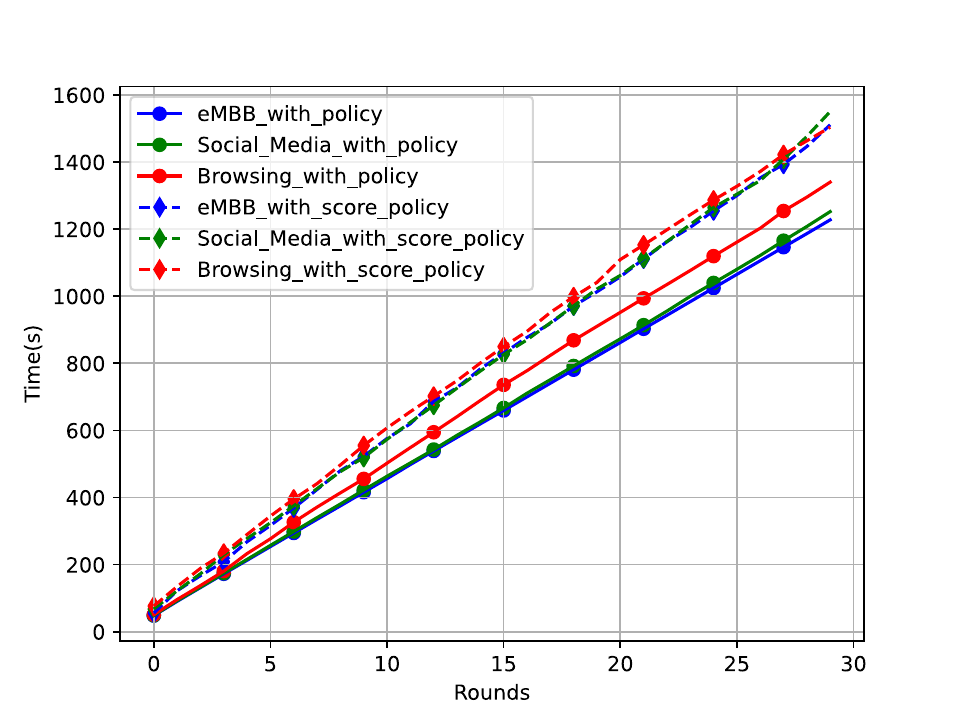}}
\caption{Completion time vs. number of FL rounds for $m=5$ and $K=10$.}
\label{Time}
\end{figure}

\hfill\\
\textbf{Convergence:} From Fig. \ref{fig:Sol_MSE} it is clear that for all the slices, the XAI-based FL converges in fewer rounds compared to the case without any policy, since AEs with higher relevance participate more frequently in the training.
Moreover, Fig. \ref{fig:Sol_Time} shows the results for the completion time where it is possible to conclude that the solution allows to decrease the time to reach convergence, thus diminishing the computation workload more quickly.

From Fig. \ref{fig:Sol_MSE_score} it is possible to observe a slight improvement in the convergence for all the slices, which is not as obvious because both solutions tend to select the clients that have a more valuable impact on the global model. Furthermore, Fig. \ref{fig:Sol_Time_score} reveals a boost in the speed of completion of the simulation run.

\begin{figure}[htbp]
\centering
\includegraphics[scale=0.5]{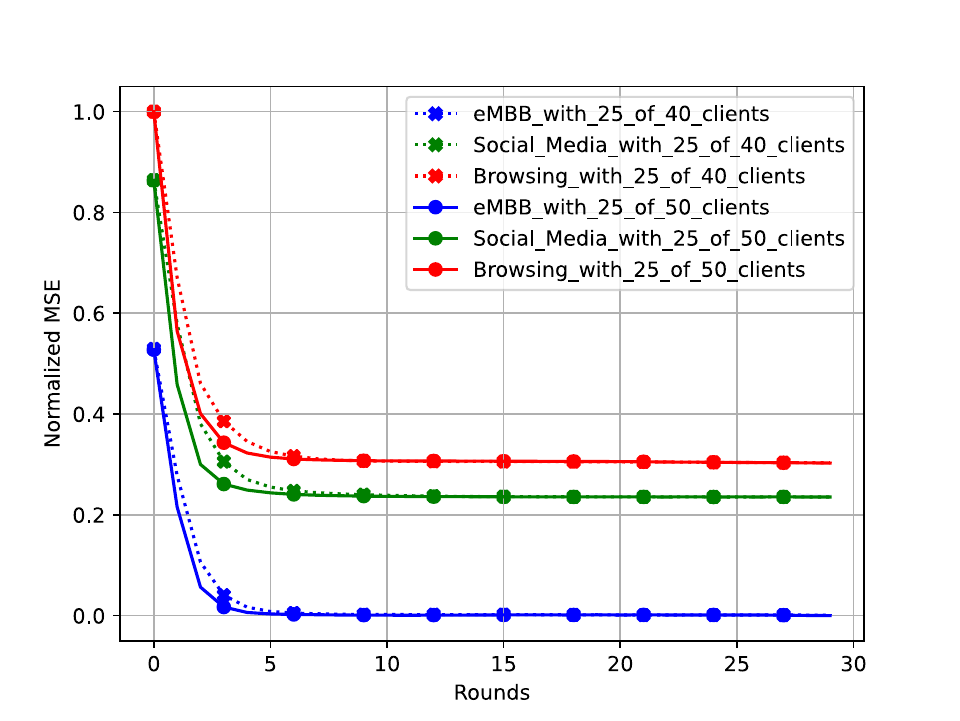}
\caption{\label{fig:Sol_MSE_40_50}FL training MSE loss vs. number of FL rounds with proposed policy for $m=25$ and $K=(40,50)$.}
\end{figure}

\textbf{Scalability:} Fig. \ref{fig:Sol_MSE_40_50} presents the results for two configurations with total AEs $K=(40,50)$, while keeping constant the number of selected AEs constant $m=25$, per round from the applied policy. It is apparent that increasing the number of total AEs does not significantly affect the performance of the solution, which ensures scalability.
That is caused by robust client selection, which ensures that selected AEs represent a useful and diverse data subset, even with a larger pool to choose from.

\begin{figure}[htbp]
\centering
\includegraphics[scale=0.54]{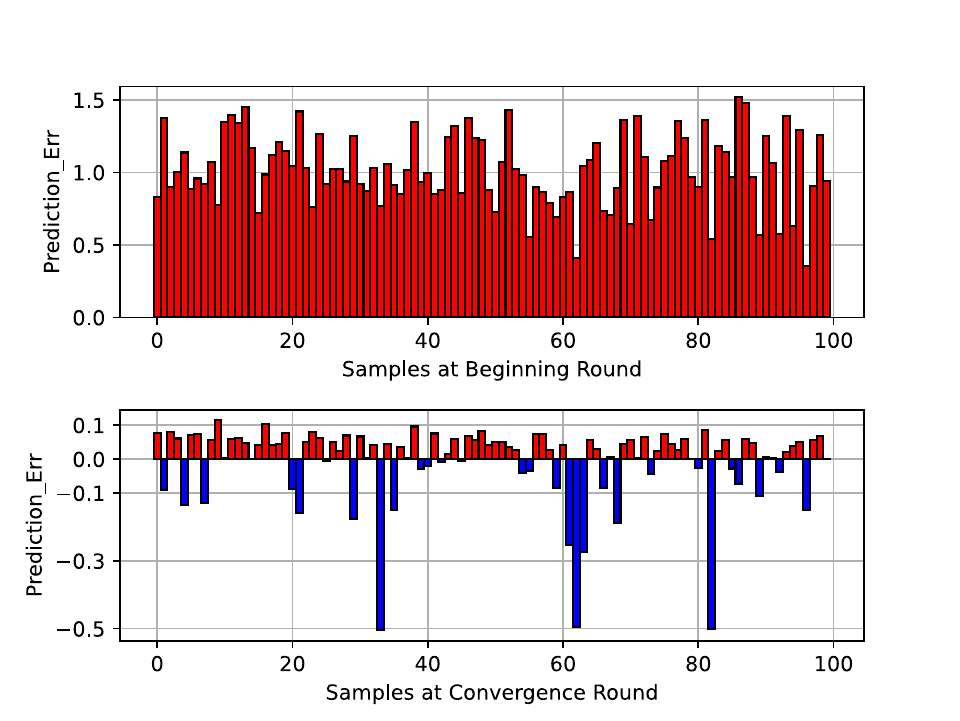}
\caption{\label{fig:Sol_provisioning}Trade-off between over- and under- provisioning for the eMBB slice of proposed policy for $m=5$ and $K=10$.}
\end{figure}

\textbf{Provisioning:} Fig. \ref{fig:Sol_provisioning} depicts the trade-off between the over- and under-resource provisioning of the model. It shows how, in the convergence round, the prediction error of a subset of samples is much less than in the initial round, underlying the good performance of the implementation, in light of the choice of the most valuable AEs during the learning process.

The prediction error is the difference between the predicted error and the actual output:
\setlength{\abovedisplayskip}{2pt}
\setlength{\belowdisplayskip}{2pt}
\begin{equation}
P_{err}(x,y)=\hat{y}_{k}^{(i)}-y_{k}^{(i)}
\end{equation}

In the plot the red part represents where the predictions surpass the real needs, and it is called over-provisioning; while, the blue part indicates less resources assigned, resulting in an under-provisioning. The presented model manages to effectively balance the trade-off, reducing in an absolute sense both over- and under- provisioning at the FL convergence.

This test was presented only for the eMBB slice, since the other slices have comparable results.

\begin{figure}[htbp]
\centering
\includegraphics[scale=0.5]{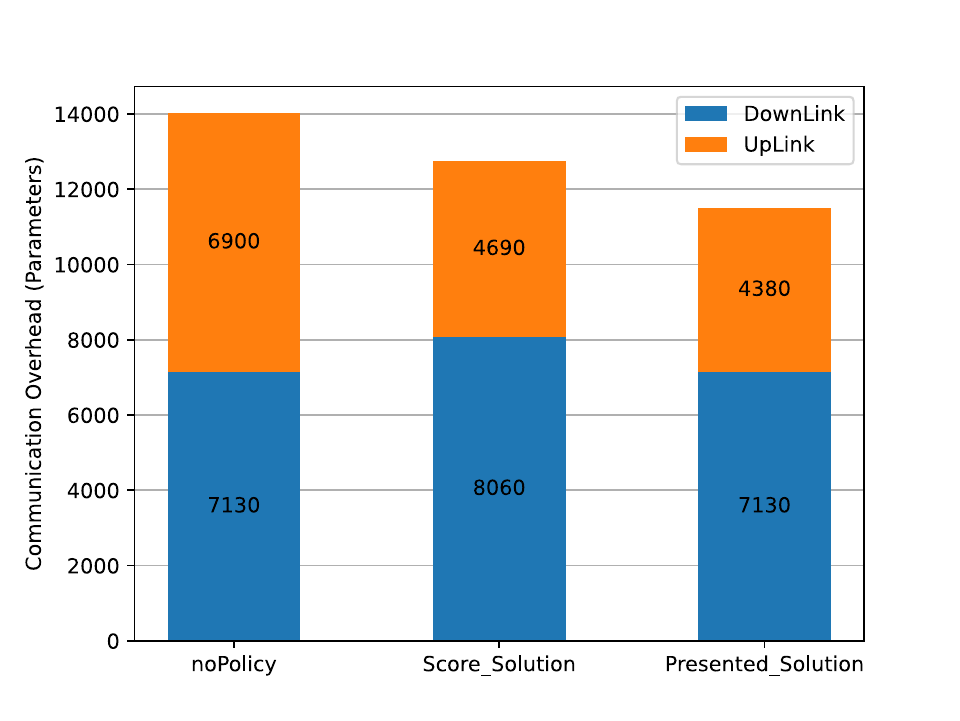}
\caption{\label{fig:Comm_Load} Performance evaluation comparison in terms of generated communication overhead. The parameters considered are: NN model's weights (in all solutions), clients attribute per feature (in Score\_Solution and Proposed\_Solution) and the clients' score (only in Score\_Solution). Scenario of 1 slice with 30 rounds for $m=5$ and $K=10$.}
\end{figure}

\textbf{Communication load:} In Fig. \ref{fig:Comm_Load}, since not all AEs are selected in every round in the proposed solution compared to the no-policy solution, the load drops in each communication iteration, attaining an overall reduction in communication overhead.
The parameters that influence the most the results are the NN parameters and the number of selected AEs with respect to the total number (the parameters are considered as single-float number sent on the link).

Observing the Score Solution column, it has a higher overhead compared to the presented solution due to the fact that it requires exchanging both the calculated global attribution weights and each evaluated AE's score for the client selection phase at each iteration.

\section{Conclusions}
In this paper, a scalable IntelliSelect-FL policy for zero-touch network slicing resource allocation has been presented. Through the selection process of a subset of active AE based on the attributions of the XAI method, the FL training achieved significant results: such as minimizing the number of rounds to reach convergence, as well as the computational time; while also reducing the communication overhead. Furthermore, the simulation results regarding the usage of different number of total AEs demonstrate that the proposed solution is scalable.
Another relevant improvement is the advantageous balance between resource over and under provisioning in fewer rounds with respect to the no-policy solution, leading to an improved overall performance. Finally, results comparison with "Score Solution" shows that the proposed solution is faster at converging and takes less time to complete the whole process.

Moreover, future research will investigate cloud-native solutions that use lightweight AI models and Kubernetes-based orchestration to enable scalable XAI-driven client selection in FL. 
This includes conducting preliminary real-world testing of FL frameworks deployed on edge-cloud infrastructures to assess the effectiveness of the proposed solution under dynamic network conditions.
Subsequently, we aim to integrate our approach into operational beyond 5G/6G testbeds, collaborating with industry partners to evaluate performance in live network environments. These deployments will help refine the model’s ability to handle practical challenges and real-time resource constraints, ensuring its readiness for large-scale adoption in next-generation wireless networks.


\section{Acknowledgments}
This research has been partially supported by the HORSE project (Grant No. 101096342), the European Union under NextGenerationEU (PRIN 2022, CUP E53D23000760006, ID 2022MWBFEE001), and the projects 6G-BRICKS (101096954) and 6G-Intense (101139266) under the Horizon Program.


\begin{thebibliography}{00}



\bibitem{5g_slice_survey} X. Foukas, G. Patounas et al., “Network slicing in 5g: Survey and challenges,” IEEE Communications Magazine, vol. 55, no. 5, pp. 94–100, 2017

\bibitem{ETSI_ZSM} ETSI GS ZSM 002, “Zero-touch network and Service Management
(ZSM); Reference Architecture,” Aug. 2019.

\bibitem{net_slic_FL_approach} B. Brik and A. Ksentini, "On Predicting Service-oriented Network Slices Performances in 5G: A Federated Learning Approach," 2020 IEEE 45th Conference on Local Computer Networks (LCN), Sydney, NSW, Australia, 2020, pp. 164-171, doi: 10.1109/LCN48667.2020.9314849

\bibitem{FL_prob} P. Kairouz et al., “Advances and Open Problems in Federated Learning,” Online. Available: arxiv.org/abs/1912.04977.

\bibitem{CL_in_FL_princ} Fu, L.; Zhang, H.; Gao, G.; Wang, H.; Zhang, M.; Liu, X. Client selection in federated learning: Principles, challenges, and opportunities. arXiv 2022, arXiv:2211.01549.

\bibitem{eff_FL_part_select} F. Lai, X. Zhu, H. V. Madhyastha, and M. Chowdhury. Oort: Efficient federated learning via guided participant selection. In USENIX Symposium on Operating Systems Design and Implementation (OSDI), 2021.

\bibitem{CL_frame_eff_FL} C. Li, X. Zeng, M. Zhang, and Z. Cao. Pyramidfl: A fine-grained client selection framework for efficient federated learning. In Conference on Mobile Computing and Networking (MobiCom), 2022.”] 

\bibitem{Fast_conv_FL} H. T. Nguyen, V. Sehwag, S. Hosseinalipour, et al. Fast-convergent federated learning. IEEE Journal on Selected Areas in Communications, pages 201–218, 2021.

\bibitem{AI_interpret_myth} Lipton, Zachary C. "The Mythos of Model Interpretability." arXiv e-prints (2016): arXiv-1606.

\bibitem{Axiomatic} M. Sundararajan, A. Taly, and Q. Yan, “Axiomatic attribution for deep networks,” in Proceedings of the 34th International Conference on Machine Learning - Volume 70, ser. ICML’17. JMLR.org, 2017, p. 3319–3328.

\bibitem{Fed_XAI} J.L.C. Bárcena, M. Daole, P. Ducange, F. Marcelloni, A. Renda, F. Ruffini, A. Schiavo, Fed-XAI: Federated learning of explainable artificial intelligence models, in: XAI.It 2022: 3rd Italian Workshop on Explainable Artificial Intelligence, Co-Located with AI*IA 2022, 2022

\bibitem{FL_nonIID_data} Yue Zhao, Meng Li, Liangzhen Lai, Naveen Suda, Damon Civin, and Vikas Chandra. 2018. Federated learning with non-IID data. arXiv:1806.00582.

\bibitem{CL_FL_pow} Cho, Y.J.; Wang, J.; Joshi, G. Client Selection in Federated Learning: Convergence Analysis and Power-of-Choice Selection Strategies. arXiv 2020, arXiv:2010.01243.

\bibitem{CL_FL_review} C. Smestad and J. Li, "A systematic literature review on client selection in federated learning", Proc. 27th Int. Conf. Eval. Assess. Softw. Eng., pp. 2-11, 2023.

\bibitem{XAI_FL_fraud_detect} T. Awosika, R. Mani Shukla and B. Pranggono, "Transparency and privacy: The role of explainable AI and federated learning in financial fraud detection", arXiv:2312.13334, 2023.

\bibitem{MEC_FL_slice} R. Ou, D. Ayepah–Mensah and G. Liu, "MEC-enabled Federated Learning for Network Slicing," 2022 International Conference on Computing, Communication, Perception and Quantum Technology (CCPQT), Xiamen, China, 2022, pp. 249-254, doi: 10.1109/CCPQT56151.2022.00050.

\bibitem{FedAVg} H.-B. McMahan et al., “Communication-Efficient Learning of Deep Networks from Decentralized Data.”, in the 20th International Conference on Artificial Intelligence and Statistics (AISTATS’2017).

\bibitem{SLA_FL_net_slic} S. Roy, H. Chergui, L. Sanabria-Russo and C. Verikoukis, "A Cloud Native SLA-Driven Stochastic Federated Learning Policy for 6G Zero-Touch Network Slicing," ICC 2022 - IEEE International Conference on Communications, Seoul, Korea, Republic of, 2022, pp. 4269-4274, doi: 10.1109/ICC45855.2022.9838376.

\bibitem{CL_FL_Survey} A. Gouissem, Z. Chkirbene and R. Hamila, "A Comprehensive Survey on Client Selections in Federated Learning", 2023 Innovation and Technological Advances for Sustainable Development (ITAS) (ITAS2023), Feb. 2023.

\bibitem{CL_grad_import} O. Marnissi, H.E. Hammouti, and E. H. Bergou, “Client selection in federated learning based on gradients importance,” arXiv preprint arXiv:2111.11204, Nov. 2021.

\bibitem{Vehic_XAI_SHAP} G. Rjoub, J. Bentahar and O. A. Wahab, "Explainable AI-based Federated Deep Reinforcement Learning for Trusted Autonomous Driving," 2022 International Wireless Communications and Mobile Computing (IWCMC), Dubrovnik, Croatia, 2022, pp. 318-323, doi: 10.1109/IWCMC55113.2022.9824617.

\bibitem{Vehic_XAI_LIME} G. Rjoub, J. Bentahar and O. A. Wahab, "Explainable Trust-aware Selection of Autonomous Vehicles Using LIME for One-Shot Federated Learning," 2023 International Wireless Communications and Mobile Computing (IWCMC), Marrakesh, Morocco, 2023, pp. 524-529, doi: 10.1109/IWCMC58020.2023.10182876.

\bibitem{Explain_FDL_slic} S. Roy, F. Rezazadeh, H. Chergui, and C. Verikoukis, “Joint Explainability and Sensitivity-Aware Federated Deep Learning for Transparent 6G RAN Slicing,” in IEEE International Conference on Communications, 2023, pp. 1238–1243.

\bibitem{dataset-4g}H. Chergui, L. Blanco and C. Verikoukis, "Statistical Federated Learning for Beyond 5G SLA-Constrained RAN Slicing," in IEEE Transactions on Wireless Communications, vol. 21, no. 3, pp. 2066-2076, March 2022, doi: 10.1109/TWC.2021.3109377. 

\bibitem{non_IID_Survey} H. Zhu, J. Xu, S. Liu and Y. Jin, "Federated learning on non-IID data: A survey", Neurocomputing, vol. 465, pp. 371-390, 2021.

\bibitem{Adam} Kingma, D.P.; Ba, J. Adam: A Method for Stochastic Optimization. In Proceedings of the 3rd International Conference on Learning Representations (ICLR), San Diego, CA, USA, 7–9 May 2015.

\bibitem{slices} H. Chergui and C. Verikoukis, "Offline SLA-Constrained Deep Learning for 5G Networks Reliable and Dynamic End-to-End Slicing," in IEEE Journal on Selected Areas in Communications, vol. 38, no. 2, pp. 350-360, Feb. 2020, doi: 10.1109/JSAC.2019.2959186.

\bibitem{tensorFlow} “Tensorflow federated,” https://www.tensorflow.org/federated, accessed: 11/29/2019.


\end{thebibliography}
\end{document}